%% file: silvotti.tex
\documentclass[11pt,twoside]{article}
\usepackage{asp2010}
\usepackage{graphicx}
\input{silvotti_defs}  

\resetcounters

\begin{document}

\title{Planetary companions to sdB stars}
\author{Roberto Silvotti$^1$, Roy H. \O stensen$^2$, John H. Telting$^3$,
and Christophe Lovis$^4$
\affil{$^1$INAF--Osservatorio Astrofisico di Torino, via dell'Osservatorio 20,
\\ ~10025 Pino Torinese, Italy}
\affil{$^2$Instituut voor Sterrenkunde, KU Leuven, Celestijnenlaan 200D, 
\\ ~3001 Leuven, Belgium}
\affil{$^3$Nordic Optical Telescope, Apartado 474, 38700 Santa Cruz de La 
Palma, \\ ~Spain}
\affil{$^4$Observatoire de Gen\`eve, Universit\'e de Gen\`eve, 51 Ch. des 
Maillettes, \\ ~1290 Sauverny, Switzerland}}

\begin{abstract}
The formation of single sdB stars is an unresolved issue and the presence of 
close sub-stellar companions could explain the stellar envelope ejection near
the tip of the RGB, that is needed to form an sdB star.
In the last 6 years the first planets orbiting sdB stars were detected using 
different methods.
In this article we will discuss these recent discoveries, review the different
detection methods used, and present some preliminary results obtained with
Harps-N, which allows to reach an unprecedented accuracy in radial velocity
measurements.
\end{abstract}

\section{SdB planets, detection methods, and the {\bf a -- M$_{\rm P}$} 
diagram}

\begin{figure}[h]
\label{sep_vs_mass}
\centering
\includegraphics[width=9.0cm,angle=0]{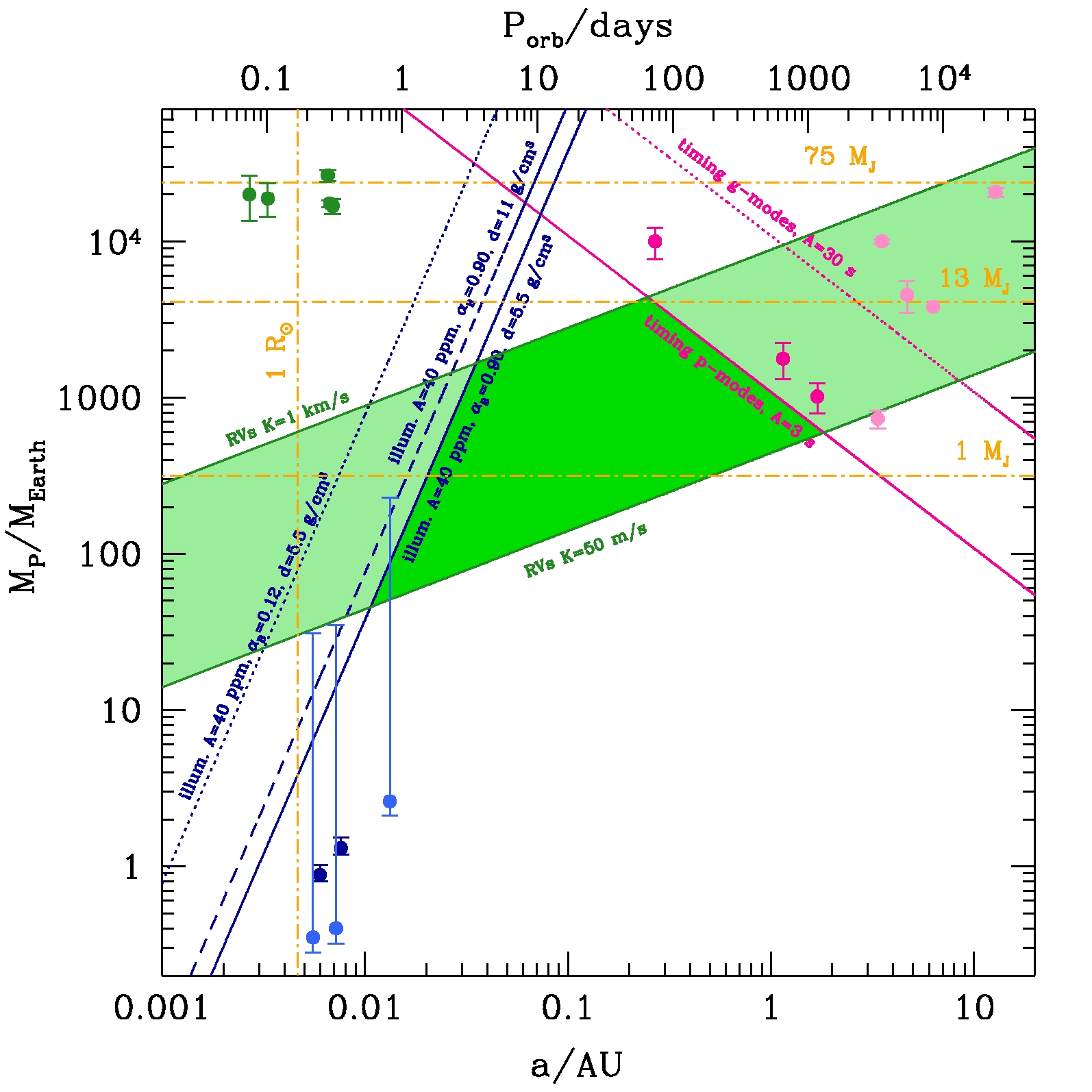}
\caption{Orbital distance vs planetary mass (or M$_{\rm P}$\,sin\,$i$ when the 
inclination is unknown).
The BD/planet candidates fall in 3 well separated regions, 
corresponding to 3 different detection methods: RVs (5 upper-left green dots); 
timing (8 upper-right dots, 3 from pulsation timing in red and 5 from eclipse
timing in pink), and illumination (5 lower-left blue dots with Earth-like 
masses detected with \kep, including three new candidates from 
\citealt{silvotti13} in cyan).
For each method current observational limits are shown: 1 km/s RVs in green, 
3 and 30 s for $p$- and $g$-mode pulsation timing in red, 
40 ppm for reflection using different Bond albedos 
($\alpha_{\rm B}$) and mean planet densities (d) in blue.
The blue lines do not include differential re-emission from heated day-side
vs cooler night-side hemisphere, which is actually the mean source of 
photometric variations for sdB stars.
This is why the blue dots are below the detection limits (see 
\citealt{charpinet11} and \citealt{silvotti13} for more details).
The green line at 50 m/s represents the limit that can be reached
with Harps/Harps-N (see Section 2). Colors available in the electronic version
only.}
%
\end{figure}

Even though only a small fraction of stars, about 2\% \citep{heber86}, become 
extreme horizontal branch stars and subdwarf B (sdB) stars in particular,
there are at least two good reasons to search for, and study, 
substellar companions to sdB stars: 
{\bf 1)} the envelope ejection that is needed to form an sdB star is well 
explained in terms of close binary evolution for half of the sdBs that have a 
close stellar companion, generally an M-dwarf or a white dwarf \citep{han02, 
han03}.
But it is more problematic for the other half of apparently single sdB stars.
The presence of close and massive planets or brown dwarfs (BDs) is a possible
explanation \citep{soker98}, that seems corroborated by recent calculations
from \citet{han12}.
Their Fig.~1 shows that, near the peak distribution of sdB masses at $\sim$0.47
\msun\ \citep{fontaine12,vangrootel13}, a low-mass sub-stellar companion 
may be sufficient to eject the common envelope (CE), even when only the orbital
energy released during the spiral-in is considered.
Adding an extra term due to thermal energy, the minimum companion mass to
eject the CE is further reduced.
{\bf 2)} The fate of close-in planets after the red giant (RG) expansion of 
their parent star is not known and observational constraints are required
to test the models.
Stellar mass loss and tidal interactions play a crucial role determining 
whether the planets migrate outwards or inwards.
In the latter case the planets transfer angular momentum to the star's 
envelope, start accreting material, and finally enter the RG envelope. 
Here their survival depends on various factors, first of all on their original 
mass.
The opposite effects of stellar mass loss and tidal interactions determine
a gap in the final distribution of orbital distances and periods
\citep{villaver09,nordhaus13}.

The first detections of substellar companions to sdB stars started six years 
ago and now we have about fourteen sdB stars with eighteen planet/BD candidates
belonging to three different groups, well distinct in terms of orbital distance
and planetary mass (Fig.~1). 
In order of decreasing orbital distance we find: 
i) eight planet/BD candidates in wide orbits (orbital periods between 3.2 and 
$\sim$16 yrs), with masses between $\sim$2 and $\sim$40 \mjup\
(\citealt{silvotti07}; \citealt{lee09} and \citealt{beuermann12b, qian09} and 
\citealt{beuermann12a, qian12}; \citealt{beuermann12a, lutz11}).
%
%
ii) Five Earth-size planet candidates, likely remnants of planetary cores,
in close orbits (orbital periods between 5.3 and 19.5 hours), detected by 
the \kep\ spacecraft \citep{charpinet11, silvotti13}.
iii) At least five BD candidates with very short orbital periods between
1.8 and 7 hours (\citealt{geier12} and references therein).
These three groups correspond to three different detection methods: timing
(both pulsation timing and eclipse timing), illumination effects, and radial
velocities (RVs) respectively, that sample three different regions of 
the {\bf a -- M$_{\rm P}$} (semi-major axis vs planetary mass) plane (Fig.~1).
More details concerning the pulsation timing and the EXOTIME program are given
by Schuh et al. (these proceedings).

\section{Pushing down the RV limits with Harps-N}

What is immediately evident from Fig.~1 is that the central region of the
{\bf a -- M$_{\rm P}$} diagram is completely unexplored.
The easiest way to access this region is to push down the present limits on 
radial velocities.
This is not an easy task considering that:\\
{\bf 1)} sdB stars are generally faint;\\
{\bf 2)} because of radiative levitation, heavy element abundances vary by 
order of magnitudes from one star to another. And a large number of metal lines
is required in order to measure low RVs (H and He lines are too broad to be of 
any utility).\\
{\bf 3)} A significant fraction of sdB stars do pulsate so that RV variations 
of $\approx$$10^2$-$10^3$ m/s can be caused by stellar pulsation.

Despite these limits, a small sample of bright (V$<$12), single,
nonpulsating, metal-rich, and slowly rotating sdB stars exists.
We recently started a program to observe them with Harps-N, an instrument
almost identical to ESO-Harps (with a $\sim$10\% higher efficiency), 
mounted at the 3.6m {\it Telescopio Nazionale Galileo} (TNG).
A few preliminary results obtained in the first run with Harps-N
are shown in Fig.~2 and 3.
These results demonstrate that, with a careful selection of the targets, 
we can measure the precise wavelength of up to 200-300 lines in the sdB spectra
and thus, through the cross correlation technique, reduce the RV errors to 
10-20 m/s, allowing to measure 50-100 m/s at 5$\sigma$ level.
In Fig.~3 we see that two targets, PG~0044+097 and PG~0342+028, show 
significant RV variations.
While PG~0342+028 was recently found to be a $g$-mode pulsator 
(E. Green, private comm.) and therefore the variations that we see are likely 
due to stellar oscillations, PG~0044+097 is a good candidate to host a 
sub-stellar companion. Further observations of this star will allow to confirm 
the presence of the companion and measure the orbital period and 
M$_{\rm P}$\,sin\,$i$. The third star shown in Fig.~3, UV~00512-08, has 
constant RVs at $\pm$30 m/s level, suggesting that the cross correlation 
technique works well.

\begin{figure}[htb]
\label{ccf}
\centering
\includegraphics[width=58mm,angle=0]{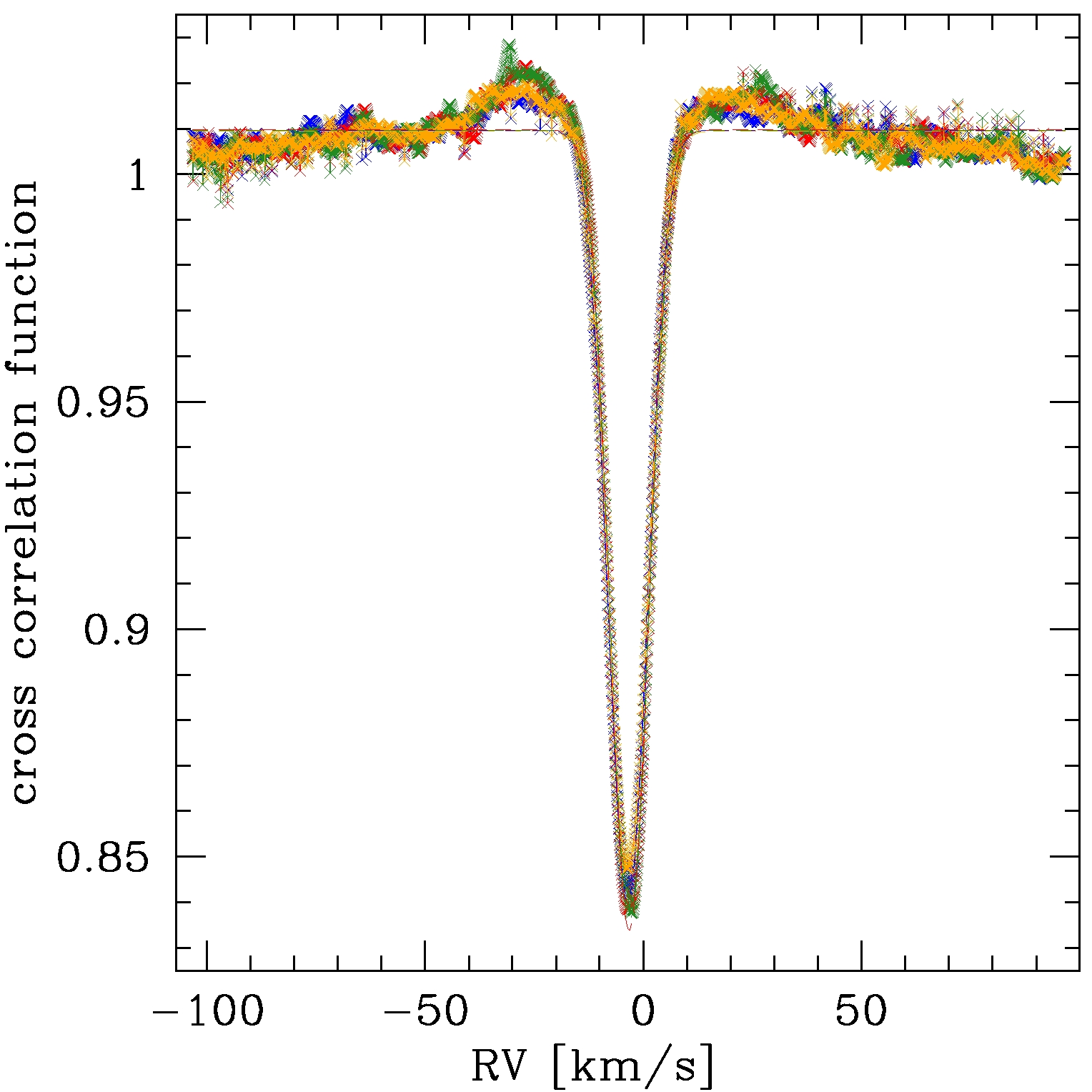}
\qquad\qquad
\includegraphics[width=58mm,angle=0]{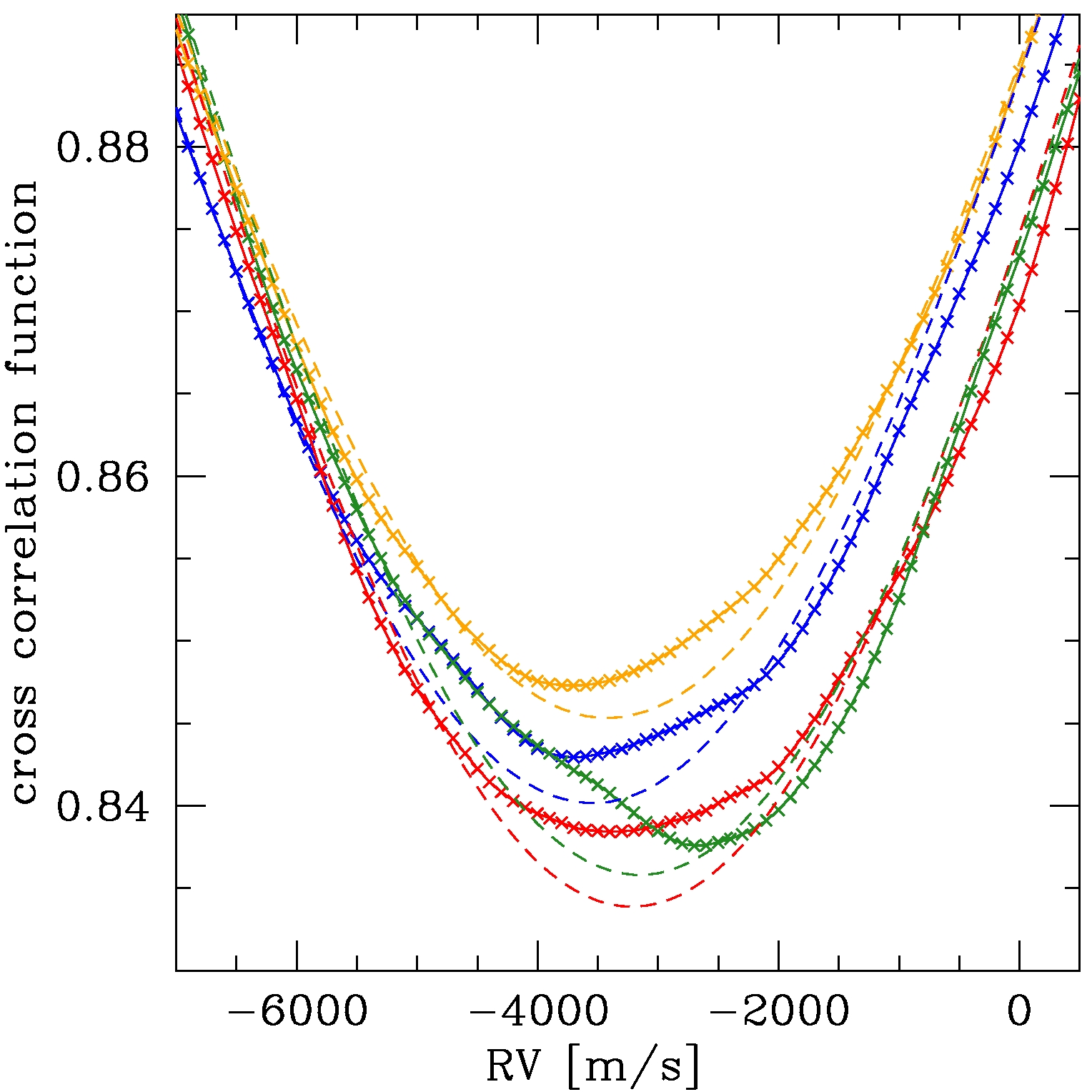}
\caption{Cross correlation function (ccf) using 212 lines from four Harps-N
spectra of PG~0044+097. 
The right plot shows in detail the ccf minima.
Each color (visible only in the electronic version) corresponds to one 
spectrum. The gaussian fits of the ccfs, from which RVs are derived, are 
represented by dashed lines.}
\end{figure}

\begin{figure}[h*]
\label{RVs}
\centering
\includegraphics[width=\textwidth,angle=0]{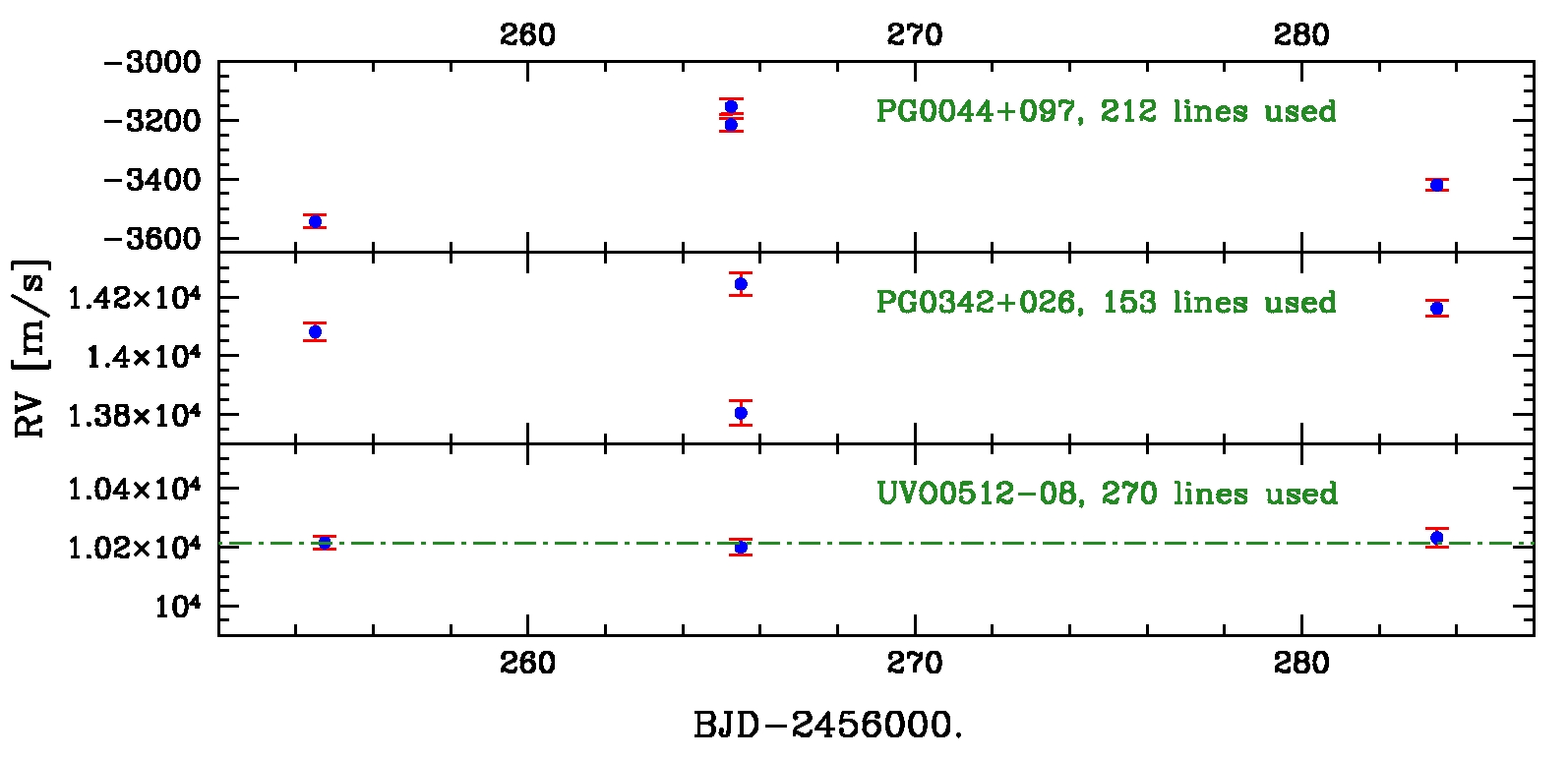}
\caption{Radial velocities for three of our targets.
See text for more details.}
\end{figure}

\section{Discussion and future}

An important step to study the sdB planets is to check whether 
the gap in orbital periods/distances predicted by different authors 
(e.g. \citealt{nordhaus13}) exists or not and which are its boundaries.
As shown in Fig.~1, the current data can not answer this question 
and we need new planet/BD candidates, in particular in the central region 
of the {\bf a -- M$_{\rm P}$} diagram, where we do not have any candidate yet.
We have seen in Section 2 that one way to start sampling the central 
region of the {\bf a -- M$_{\rm P}$} diagram is to improve the RV precision.

The reason why, apparently, the close-in planets are divided into two groups
with very different masses (Fig.~1) is another interesting aspect that could
be related only to the different detection methods used, but could also 
indicate that two different classes of objects exist that survived the engulfment: those with a sufficient mass to not have been significantly 
altered in their structure, and those with a lower mass that have been 
disrupted in one or more fragments of planetary cores.
A method to verify whether close-in substellar companions to sdBs are 
relatively common is to search for transits.
In the \kep\ sdB sample of 49 objects (including a few He sdBs), 2 show
signatures of remnants of planetary cores \citep{charpinet11, silvotti13}.
But only 15 pulsating sdBs of the \kep\ sample have been observed for a time 
sufficiently long to detect very low photometric variations of 20-50 ppm.
Assuming that 2/15 of sdBs have remnants of planetary cores at a=0.005 AU from 
the star with R$_{\rm P}<<R_{\rm sdB}$ and R$_{\rm sdB}$=0.001 AU 
(i.e. a transit probability of 0.2), the probability to observe a transit 
is about 0.027, which means that, on average, we need to observe $\sim$40 sdB 
stars to see one transit.
Considering now the more massive group of sub-stellar companions,
\citet{geier12} found that 16\% of their sample of 27 bright single-lined 
sdB stars do show small RV variations, that can be associated with massive sub-stellar companions.
But in this case the transit probability at 0.005 AU is about 0.28 (assuming 
R$_{\rm P}$=0.9 R$_{\rm Jup}$, \citealt{burrows11}, and including a $\sim$10\%
inflation of the companion, see e.g. \citealt{geier09}).
So that, on average, 1 sdB out of $\sim$20 should show transits (and secondary 
eclipses).

Finally it is worth mentioning that most of the planet/BD candidates in Fig.~1 
were found using only one detection method.
To confirm them with an independent method would be very important.
A discussion of this crucial aspect is given by Schuh et al.
(these proceedings).

\acknowledgements The authors thank the organizers for a very fruitful and 
pleasant sdB/sdO meeting.

\bibliography{silvotti.bbl}

\end{document}

%% file: silvotti_defs.tex
\newcommand{\kep}{{\em Kepler}}

\newcommand{\msun}{${\mathrm{M}}_{\odot}$}

\newcommand{\mjup}{$\rm M_{\rm Jup}$}

\def\apjl{ApJ}%
\def\aap{A\&A}%